\documentclass[1p,times,procedia]{elsarticle}
\usepackage{nupha_ecrc}
\usepackage{graphicx}	
\usepackage{xspace}

\volume{00}

\firstpage{1}

\journalname{Nuclear Physics A}

\runauth{Prakhar Garg (for the PHENIX Collaboration)}

\jid{nupha}

\jnltitlelogo{Nuclear Physics A}

\usepackage{amssymb}
 \usepackage{amsthm}

\usepackage[figuresright]{rotating}
\usepackage{graphics}

\newcommand{\sqsn}{\mbox{$\sqrt{s_{_{NN}}}$}\xspace}
\newcommand{\auau}{\mbox{Au$+$Au}\xspace}
\newcommand{\pt}{\mbox{$p_T$}\xspace}
\begin{document}
\begin{frontmatter}





\title{PHENIX results on fluctuations and Bose-Einstein
correlations in Au+Au collisions from the RHIC
Beam Energy Scan}

\author{Prakhar Garg (for the PHENIX Collaboration)\footnote{For the full PHENIX Collaboration author list and acknowledgments,
see Appendix ``Collaboration'' of this volume.}}
\address{Department of Physics, Banaras Hindu University, Varanasi 221005, India}

\begin{abstract}

The RHIC Beam Energy Scan focuses on mapping the QCD phase diagram and pinpointing the
location of a possible critical end point. Bose-Einstein correlations and event-by-event fluctuations
of conserved quantities, measured as a function of centrality and collision energy, are promising
tools in these studies. Recent lattice QCD and statistical thermal model calculations predict that
higher-order cumulants of the fluctuations are sensitive indicators of the phase transition. Products
of these cumulants can be used to extract the freeze-out parameters~\cite{Bazavov:2012vg} and to locate the critical point~\cite{Stephanov:1999zu}.
Two-pion interferometry measurements are predicted to be sensitive to potential softening of the
equation of state and prolonged emission duration close to the critical point~\cite{Pratt:1984su}. We present recent
PHENIX results on fluctuations of net-charge using high-order cumulants and their
products in Au+Au collisions at \sqsn = 7.7 - 200 GeV, and measurement of two-pion 
correlation functions and emission-source radii in Cu+Cu and Au+Au collisions at several beam energies. The
extracted source radii are compared to previous measurements at RHIC and LHC in order to study energy
dependence of the specific quantities sensitive to expansion velocity and emission duration. Implications for the
search of a critical point and baryon chemical potentials at various collision energies are discussed.

\end{abstract}


\end{frontmatter}


\section{Introduction}
\label{introduction}

Quantum chromodynamics (QCD) predicts a phase transition from a hadron gas (HG) phase to a quark gluon plasma (QGP) phase  with variations of thermodynamic parameters such as temperature ($T$) and/or baryon density ($\mu_{B}$)~\cite{Fodor:2004nz}. 
Lattice QCD calculations indicate that the chiral and de-confinement phase transitions are a smooth crossover along the temperature axis, i.e. with $\mu_{B}$ = 0, while various other models predict that the phase transition becomes first order at high baryon density~\cite{Halasz:1998qr}. The existence of the QCD critical point is thus expected as the first order phase transition line should end somewhere at finite $\mu_{B}$ and $T$. 
In order to study the properties of QGP in these experiments, it is important to choose an observable which is sensitive enough to the medium property in the early stage.

It has been proposed that the shapes of the event-by-event net-charge distributions are sensitive to
the presence of the critical point, as they are related to the conserved number susceptibilities of the system and hence to the correlation length~\cite{Gavai:2010zn}. Additionally, the shape of the emission source function can also provide signals for a second-order phase transition or proximity to the QCD critical point~\cite{Csorgo:2005it}. Two-pion correlation measurements provide important information about the space-time evolution of the particle emitting source in the collision. An emitting system which undergoes a strong first order phase transition is expected to demonstrate a much larger space-time extent than would be expected if the system had remained in the hadronic phase throughout the collision process.

The PHENIX detector at RHIC has explored the above possibilities in the recent Beam Energy Scan (BES) program of RHIC. During 2010 and 2011, RHIC provided \auau collisions to PHENIX at \sqsn = 200 GeV, 62.4 GeV, 39 GeV, 27 GeV, 19.6 GeV, and 7.7 GeV. PHENIX recorded Cu+Cu collisions at \sqsn = 200 GeV during 2005. 
Results from PHENIX covering net-charge fluctuations and two-pion interferometry measurements, are discussed here.

\section{Net-charge Fluctuations}

\begin{figure}[ht!]
 \begin{center}
\includegraphics[scale=0.30]{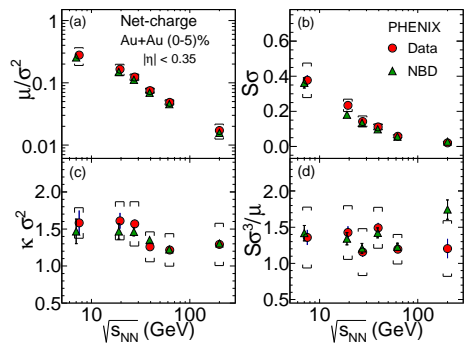}
 \caption{(Color online) The ratios of cumulants of net-charge distributions (a) $\mu/\sigma^{2}$ (b) S$\sigma$ (c) $\kappa\sigma^{2}$, and (d) S$\sigma^{3}/\mu$, after efficiency corrections for most central (0-5\%) \auau collisions. The statistical and systematic errors are shown by bars and caps, respectively. Triangles represent the efficiency corrected cumulant ratios extracted from NBD fits to positively and negatively charged particles distributions~\cite{Adare:2015aqk}.}
 \label{fig1}
 \end{center}
 \end{figure}

PHENIX has measured the distributions of net-charge multiplicity (N = $N^{+}$ - $N^{-}$) and their various moments (mean ($\mu$) =${<N>}$, variance ($\sigma^2$) = ${<(N-\mu)^2>}$, skewness (S) = $\frac{<(N-\mu)^3>}{\sigma^3}$ and kurtosis ($\kappa$) =$\frac{<(N-\mu)^4>}{\sigma^4}  -3$ ) at several beam energies~\cite{Adare:2015aqk}. The charged hadrons selected for this analysis cover transverse momentum (\pt) between 0.3 and 2.0 GeV/$c$ and pseudorapidity range spanning $|\eta|\leq  0.35$. 
Figure~\ref{fig1} shows the efficiency corrected $\mu/\sigma^2$, $S\sigma$, $\kappa\sigma^2$, and $S\sigma^3/\mu$ as a function of \sqsn for the most central (0-5\%) \auau collisions. 
In Fig.~\ref{fig1}, triangles represent the efficiency corrected cumulants ratios extracted from NBD fits to positively and negatively charged particles distributions. The $\kappa\sigma^{2}$ values are positive and constant at all the collision energies within the statistical and systematic uncertainties as is shown in Fig.~\ref{fig1}.

\begin{figure}[ht!]
 \begin{center}
\includegraphics[scale=0.27]{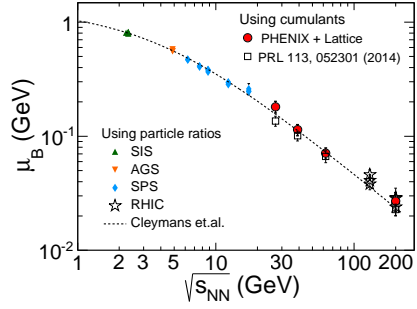}
 \caption{(Color online) Chemical freeze-out parameter ($\mu_{B}$), extracted from PHENIX higher-moments analysis, as a function of center of mass energy (\sqsn) are shown in red solid points~\cite{Adare:2015aqk}. The dashed line shows the parametrization given in Ref.~~\cite{Cleymans:2005xv} and the other experimental data are from Ref.~\cite{Cleymans:2005xv} and references therein.}
 \label{fig2}
 \end{center}
 \end{figure}

Comparing these measurements with the lattice calculations, freeze-out temperature ($T_{f}$) and baryon chemical potentials ($\mu_{B}$) are also extracted at freeze-out. Figure~\ref{fig2} shows the variation of $\mu_{B}$ as a function of \sqsn. The extracted $\mu_{B}$ values are found comparable to the $\mu_{B}$ values extracted from particle ratio analysis given in Ref.~\cite{Cleymans:2005xv}.


\section{Two-pion interferometry}
\begin{figure}[ht!]
 \begin{center}
\includegraphics[scale=0.32]{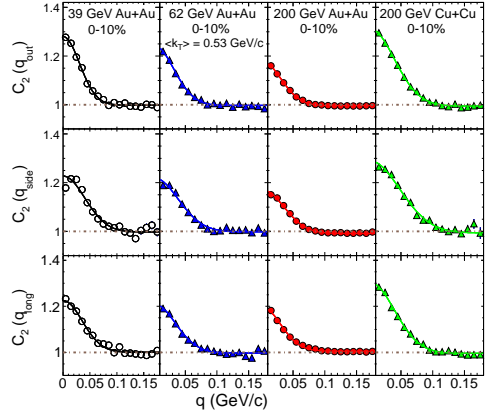}
\caption{(Color online) Correlation functions of two-pion pairs ($\pi^{+}\pi^{+}$ and $\pi^{-}\pi^{-}$) for 0-10\% central \auau (left) and Cu+Cu (right) collisions for pion pair transverse momenta ($\langle{k_T}\rangle$) = 0.53 GeV/c and for several \sqsn. The curves represent fits to the correlation function~\cite{Adare:2014qvs}.  
}
 \label{fig3}
 \end{center}
 \end{figure}

\begin{figure}[ht!]
 \begin{center}
\includegraphics[scale=0.40]{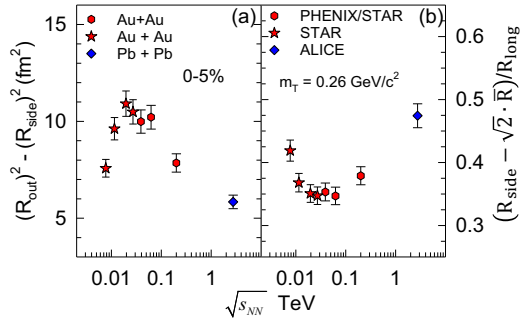}
 \caption{(Color online) 
 The $\sqsn$ dependence of (a) ($R^{2}_{out} - R^{2}_{side}$), (b) [({$R_{side}$ -$\sqrt{2}\bar{R})/R_{long}$}].   The PHENIX and STAR data points represent the results from fits to the $m_{T}$ dependence of the combined data sets~\cite{Adare:2014qvs}.
 }
 \label{fig4}
 \end{center}
 \end{figure}

PHENIX has performed measurements of two-pion correlations in Cu+Cu collisions at \sqsn = 200 GeV and Au+Au collisions at \sqsn = 39, 62.4 and 200 GeV~\cite{Adare:2014qvs}. 
Figure~\ref{fig3} shows the two-pion correlation functions as a function of the components of the momentum difference ($\bf{q}$) between particles in the pair for several \sqsn. These correlation functions are fitted with a function which incorporates Bose-Einstein enhancement and the Coulomb interaction between the pairs, to extract the HBT radii ($R_{side}$, $R_{out}$ and $R_{long}$ ). The quantities, $R^{2}_{out} - R^{2}_{side}$ and $(R_{side} - \sqrt{2}\bar{R})/R_{long}$ (see reference~\cite{Bhalerao:2005mm} for $\bar{R}$), which are related to emission duration and medium expansion velocity, respectively, are shown (Fig.~\ref{fig4}) for pair transverse mass $m_{T}$ = 0.26 GeV/$c^{2}$ to reduce the effect of position momentum correlation. Also, the PHENIX results are compared with STAR results for \sqsn = 7-200 GeV and ALICE results at LHC for \sqsn = 2.76 TeV. A maximum is observed as a function of \sqsn in $R^{2}_{out} - R^{2}_{side}$~(Fig.~\ref{fig4}(a)) with complimentary minimum in $(R_{side} - \sqrt{2}\bar{R})/R_{long} ($Fig.~\ref{fig4} (b)). Non-monotonic behavior over a small range in \sqsn may point to a softening of equation of state that may coincide with the QCD critical point.

\section{Summary}
PHENIX results for net-charge fluctuations and two-pion interferometry as a function of beam energy are presented. The net-charge fluctuation measurements do not give a clear indication of the presence of the QCD critical point, though the $\mu_{B}$ extracted with lattice calculations and PHENIX data are found to be consistent with previously extracted baryon chemical potentials.  
A non-monotonic behavior in the quantities related to emission duration and medium expansion velocity is observed, which hints the softening of equation of state. Further, more detailed studies are required for a clear picture of QCD phase diagram.

  \end{document}